\documentstyle[aas2pp4]{article}

\def\simless{\mathbin{\lower 3pt\hbox
     {$\rlap{\raise 5pt\hbox{$\char'074$}}\mathchar"7218$}}}   
\def\simmore{\mathbin{\lower 3pt\hbox
     {$\rlap{\raise 5pt\hbox{$\char'076$}}\mathchar"7218$}}}   

\received{}
\accepted{}
\slugcomment{Submitted to Ap.J.Letters., October 1997}

\lefthead{M.~M\'endez et al.}
\righthead{Discovery of a second kHz QPO peak in 4U~1608--52.}

\begin{document}

\title{Discovery of a second kHz QPO peak in 4U~1608--52}

\author{M.~M\'endez\altaffilmark{1,2},
        M.~van~der~Klis\altaffilmark{1},
        J.~van~Paradijs\altaffilmark{1,3},
        W.~H.~G.~Lewin\altaffilmark{4},
        B.~A.~Vaughan\altaffilmark{5},
        E.~Kuulkers\altaffilmark{6},
        W.~Zhang\altaffilmark{7},
        F.~K.~Lamb\altaffilmark{8},
        D.~Psaltis\altaffilmark{9}}

\altaffiltext{1}{Astronomical Institute ``Anton Pannekoek'',
       University of Amsterdam and Center for High-Energy Astrophysics,
       Kruislaan 403, NL-1098 SJ Amsterdam, the Netherlands}

\altaffiltext{2}{Facultad de Ciencias Astron\'omicas y Geof\'{\i}sicas, 
       Universidad Nacional de La Plata, Paseo del Bosque S/N, 
       1900 La Plata, Argentina}

\altaffiltext{3}{Physics Department, University of Alabama in Huntsville,
       Huntsville, AL 35899, USA}

\altaffiltext{4}{Massachusetts Institute of Technology, Center for Space
       Research, Room 37-627, Cambridge, MA 02139, USA}

\altaffiltext{5}{Space Radiation Laboratory, California Institute of
       Technology, MC 220-47, Pasadena CA 91125, USA}

\altaffiltext{6}{Astrophysics, University of Oxford, Nuclear and
       Astrophysics Laboratory, Keble Road, Oxford OX1 3RH, United
       Kingdom}

\altaffiltext{7}{Laboratory for High Energy Astrophysics, Goddard
       Space Flight Center, Greenbelt, MD 20771, USA}

\altaffiltext{8}{Departments of Physics and Astronomy, University of
       Illinois at Urbana-Champaign, Urbana, IL 61801, USA}

\altaffiltext{9}{Harvard-Smithsonian Center for Astrophysics,
       60 Garden Street, Cambridge, MA 02138, USA}

\begin{abstract}

Using a new technique to improve the sensitivity to weak Quasi-Periodic
Oscillations (QPO) we discovered a new QPO peak at about 1100~Hz in the
March 1996 outburst observations of 4U~1608--52, simultaneous with the
$\sim 600 - 900$~Hz peak previously reported from these data.  The
frequency separation between the upper and the lower QPO peak varied
significantly from $232.7 \pm 11.5$~Hz on March 3, to $293.1 \pm 6.6$~Hz
on March 6.  This is the first case of a variable kHz peak separation in
an atoll source.  We discuss to what extent this result could be
accommodated in beat-frequency models such as proposed for the kHz QPOs.
We measured the rms fractional amplitude of both QPOs as a function of
energy, and we found that the relation is steeper for the lower than for
the upper frequency peak.  This is the first source where such
difference between the energy spectrum of the two kHz QPOs could be
measured.

\end{abstract}


\section{Introduction}

Recent observations with the Rossi X-ray Timing Explorer (RXTE) have led
to the discovery of kilohertz quasi-periodic oscillations (kHz QPOs) in
about a dozen low-mass X-ray binaries (LMXB; see van der Klis
\cite{vanderklis97a} for a review).  In most cases the power spectra of
these sources show twin kHz peaks that move up and down in frequency
together, keeping a constant separation (e.g., Strohmayer et al.
\cite{strohmayer96a}, Wijnands et al.  \cite{wijnandsetal97}, Ford et
al.  \cite{ford97a}).  Sometimes a third kHz peak is detected near a
frequency equal to the separation frequency of the twin peaks
(Strohmayer et al.  \cite{strohmayer96a}) or twice that (Wijnands \& van
der Klis \cite{wijnands97}; Smith, Morgan, \& Bradt \cite{smith97}),
indicating a beat-frequency interpretation, with the third peak near the
neutron star spin frequency (or twice that value) (Strohmayer et al.
\cite{strohmayer96b}).  However, in Sco~X--1 the twin peak separation
varies, which is not consistent with a simple beat-frequency
interpretation (van der Klis et al.  \cite{vanderklisetal97a}).

The transient atoll source (Hasinger \& van der Klis \cite{hasinger89})
4U~1608--52 was observed with RXTE on 1996 March 3, 6, 9, and 12 during
an outburst (Berger et al.  \cite{berger96}).  In the first three
pointings a single QPO peak was detected in the power spectra, with a
frequency that changed from $\sim 570$ to $\sim 890$~Hz, while on March
12 no QPO peak was detected at frequencies $\simmore 100$~Hz.  After the
outburst, further observations were performed by RXTE on 1996 August 1,
October 24, and 1997 July 18.

In this paper we report on the results obtained from a reanalysis of the
data discussed by Berger et al.  (\cite{berger96}), including more data
from March 6 not analyzed in that paper, and an analysis of the new
observations of 4U~1608--52 carried out by RXTE since then.  Using a new
technique to increase the sensitivity to weak QPOs, on March 3 and 6 we
detected a second peak near 1100~Hz, simultaneous with the $\sim 600 -
900$~Hz peak, which went previously unnoticed.

\section{Observations and Data Analysis}

We describe four RXTE observations of 4U~1608--52 in outburst on 1996
March 3, 6, 9, and 12 (see Berger et al.  \cite{berger96}), plus three
additional ones spanning 3, 3.8, and 6~ksec starting UTC 1996 August 1
12:13:20, 1996 October 24 22:30:24, and 1997 July 18 02:04:16,
respectively.

High time resolution data were collected on March 6 only from channels 0
to 49 ($\sim 2 - 12.7$~keV), and at all other times in the $2 - 60$~keV
range.  In all cases there were simultaneous 16~s time resolution data
in 129 bands covering the full energy range of the PCA.

We used the 16~s data to produce X-ray spectra for all the observations,
which we fitted with a blackbody plus a power law model (see Ford et al.
\cite{ford97b} for a detailed analysis of the the energy spectra of
these observations).  From these fits, we find that the unabsorbed 2 -
20~keV fluxes on 1996 March 3, 6, 9, 12, August 1, October 24, and 1997
July 18 were 10.56, 7.54, 1.62, 1.61, 1.04, 0.65, and 0.04 $10^{-9}$
erg~cm$^{-2}$~s$^{-1}$, respectively, corresponding to full PCA source
count rates (background subtracted) of $\sim 2785 - 3265$~c/s, $\sim
1980 - 2480$~c/s, $\sim 640 - 670$~c/s, $\sim 620 - 780$~c/s, $\sim 370
- 410$~c/s, $\sim 210 - 260$~c/s, and $\sim 5 - 50$~c/s, respectively
(notice that the count rates quoted by Berger et al.  \cite{berger96}
for the March 6 observation were for $6 - 12.7$~keV).  We observed no
X-ray bursts.

For each individual observation we calculated power spectra using 64~s
data segments, and averaged them.  We renormalized all power spectra to
fractional rms squared per Hertz (see van der Klis \cite{vanderklis95})
using background measurements from slew and Earth occultation data.

On March 3 and 6 the very strong peak previously reported by Berger et
al.  (\cite{berger96}) was obvious.  Its frequency varied from $\sim
820$ to $\sim 890$~Hz, (see Figs.  1 and 2 of Berger et al.
{\cite{berger96}) on March 3, and between $\sim 650$ and $\sim 870$~Hz
on March 6, extending the range seen by Berger et al.  (\cite{berger96})
by $\sim 200$~Hz due to our current more complete data set.  Close
inspection of these power spectra suggested to us the presence of a
second QPO at a frequency $\sim 200 - 300$~Hz higher than that of the
strong peak.  However, in both cases this peak was very broad, and only
marginally significant.  Scrutiny of a dynamical power spectrum of March
6 further suggested that the second peak matched the frequency
variations of the first one.

In order to investigate the reality of this second peak we divided the
data in segments of 64~sec and calculated a power spectrum for each
segment.  The strong peak was well detected in each segment.  We then
fitted the centroid frequency of the strong peak in each individual
power spectrum, and shifted the frequency scale of each spectrum to a
frame of reference where the position of the strong peak was constant in
time.  Finally, we averaged these shifted power spectra.

If the frequency separation between the two peaks were constant, then
this ``shift and average'' procedure to compensate for the frequency
change of the strong peak would also compensate for the frequency change
of the weak peak, optimizing chances to detect it.  The improvement in
the sensitivity comes from the fact that the signal-to-noise ratio $S/N$
of a QPO peak of given rms amplitude is inversely proportional to the
square root of its width (van der Klis \cite{vanderklis89}), and the
motion of the peak, if uncorrected, makes it much wider, reducing $S/N$.

\section{Results}

In Fig.~\ref{psfig} we show the power spectra of 1996 March 3 and 6
calculated using the above method.  In both power spectra a second QPO
peak can be seen at a frequency $\sim 200 - 300$~Hz above that of the
strong peak of Berger et al.  (\cite{berger96}).  We fitted the $256 -
3750$~Hz range of each power spectrum with a function consisting of a
constant level, representing the Poisson noise, the Very Large Event
contribution (Zhang et al.  \cite{zhang95i}; Zhang \cite{zhang95ii}),
and two Lorentzians.  On March 6 we also included a power law to account
for the broad-band noise component at low frequencies.  The results are
in Table~\ref{tabpow}; we do not quote the peaks' centroid frequencies,
which were arbitrarily shifted, but only the peak separation, $\Delta
\nu$.  Frequencies of the lower peak are reported below
(Fig.~\ref{freq_rate}).  The $1\sigma$ error bars from the fits indicate
the second peak to be $4.3 \sigma$ and $4.4 \sigma$ significant on March
3, and March 6, respectively.  An $F$-test to the $\chi^{2}$ of the fits
with and without this peak yields a probability of $7.1 \times 10^{-10}$
on March 3, and $5.3 \times 10^{-8}$ on March 6, for the null hypothesis
that the peak is not present in the data.  Considering the number of
trials implied by the number of independent frequencies analyzed (van
der Klis \cite{vanderklis89}), these probabilities increase to $1.4
\times 10^{-7}$ and $1.1 \times 10^{-5}$, respectively.

Interestingly, $\Delta \nu$ changed from $232.7 \pm 11.5$~Hz on March 3
to $293.1 \pm 6.6$~Hz on March 6, a change of $60.4 \pm 13.3$~Hz.  We
tested the significance of this result by fitting both power spectra
simultaneously, but forcing the distance between the peaks to be the
same in both of them.  Applying an $F$-test to the $\chi^{2}$ of this
fit and the fit where all parameters were free we get a probability of
$2.4 \times 10^{-3}$ for the hypothesis that the peak separation did not
change between the two observations:  the difference in the frequency
separation between March 3 and 6 is significant at the $3.2\sigma$
level.

The FWHM of the low frequency peak in the shifted and averaged power
spectrum is 4.8~Hz on March 3, and 4.7~Hz on March 6.  The high
frequency QPO peak was broader.  It did not vary significantly between
the two observations (Table \ref{tabpow}).  The only other source where
the upper peak was measured to be significantly broader than the lower
one is 4U~1728--34 (Strohmayer et al.  \cite{strohmayer96a}).  Although
the upper QPO might be intrinsically broader, it might also have been
blurred in the shift process due to small changes in the frequency
separation during each observation.  We tried to test this by dividing
the data into shorter segments, but the lower statistics prevented us
from getting a significant result.

On March 9 we only detected a single QPO, moving from $\sim 570$ to
$\sim 735$~Hz, with upper limits to the rms of a second QPO below the
March 3 and 6 values (Table \ref{tabpow}).

On 1996 March 12, August 1, and October 24, and 1997 July 18 we detected
no QPO above 100~Hz.  The 95\,\% confidence upper limits for a $\sim
10$~Hz FWHM peak were 4.1\,\%, 7.0\,\%, 7.4\,\% and 7.8\,\% rms,
respectively, and about twice those values for a $\sim 200$~Hz FWHM
peak.  On March 12 the flux was similar to that on March 9, but no QPOs
were detected, implying that the amplitude of the QPO was $\sim 3$ times
smaller, or its FWHM was much broader, on March 12 than on March 9.
This shows that the properties of the kHz QPOs do not depend only on
flux.

In Fig.~\ref{freq_rate} we plot QPO frequency versus count rate for 1996
March 3, 6, and 9.  We also plot results of March 15 and 22 (Yu et al.
\cite{yu97}).  In nearly all cases frequency is positively correlated to
the source count rate.  On March 3 it remains more or less constant
below 3090 c/s, and then increases linearly above that value, while on
March 6 there is a linear relation (a correlation in the data of March 6
was already reported by Berger et al.  \cite{berger96}).  On March 9
there is a marginal indication of an anticorrelation of frequency with
count rate.  On March 15 and 22, at lower count rates, the frequency
seems to be correlated with count rate (Yu et al.  \cite{yu97}).

We also measured the photon energy dependence of the new higher
frequency QPO.  On March 3 we divided the data into four energy bands, 2
-- 4 -- 6 -- 8.1 -- 31.8~keV, and found rms fractional amplitudes of $<
2\,\%, 2.4 \pm 1.0\,\%, 6.1 \pm 0.8\,\%,$ and $4.5 \pm 1.0\,\%$,
respectively.  On March 6, due to the constraints of the observing mode,
we could only divide the data into two energy bands, 2 -- 6 -- 12.7~keV.
The rms fractional amplitudes were then $2.1 \pm 0.5\,\%$ and $3.8 \pm
0.5\,\%$, respectively, consistent with those of March 3.  The rms
amplitude of the high frequency QPO seems to increase with energy, but
significantly less steeply than in the low frequency QPO (Berger et al.
\cite{berger96}).  On March 9, for the same energy bands as on March 3,
the rms amplitudes of the only QPO seen in the data were $12.9 \pm
1.4\,\%, 11.5 \pm 1.4\,\%, 18.4 \pm 1.6\,\%$, and $21.3 \pm 1.3\,\%$,
compatible with both the upper and lower peak on March 3 and 6.  This
does not allow us to establish which peak we detected on March 9.

\section{Discussion}

We have found, for the first time, the second peak in 4U~1608--52
expected on the basis of comparison to other kHz QPO sources.  Compared
to the lower frequency peak, the amplitude of the higher frequency QPO
increases less steeply with energy.  We see, for the first time in a
source with a luminosity far below Eddington, the separation between the
two peaks vary.  The dependence of the QPO frequency on count rate is
complex.

Fig.~\ref{freq_rate} shows that while for most observations the QPO
frequency is correlated to count rate, there is no simple function that
fits all the observations simultaneously.  The changes in peak
separation are only moderate, so this is true for both QPO.

A similar effect has been observed when different sources, spanning a
very large range of luminosities, are plotted together in a single
frequency-luminosity diagram:  each source shows a positive correlation,
along lines which are more or less parallel (van der Klis 1997a,b; van
der Klis et al.  \cite{vanderklisetal97b}).  A similar behavior as we
see in 4U~1608--52 was also observed in 4U~0614+091 (Ford et al.
\cite{ford97a}, M\'endez et al.  \cite{mendez97}).  The fact that this
is observed in individual sources shows that a difference in neutron
star properties such as mass or magnetic field strength can not be the
full explanation for the differences observed in the
frequency-luminosity relations.

As noted by Berger et al.  (\cite{berger96}), on March 6 frequency and
count rates are correlated --we now see that the higher count rate data
on March 3 also show a correlation.  There is therefore no evidence for
the frequency hitting a ceiling towards higher count rates, although the
frequency did not decrease much when the count rate dropped from 3000 to
2400~c/s.

We now turn to the behavior of the peak separation.  It has been
proposed that the higher frequency kHz QPO represents the Keplerian
frequency, $\nu_{\rm K}$, of the accreting material in orbit around the
neutron star at some preferred radius (van der Klis et al.
\cite{vanderklis96}), while the lower frequency peak, at $\nu_{\rm B}$,
is produced by the beating of $\nu_{\rm K}$ with another frequency,
$\nu_{\rm S}$, identified as the spin frequency of the neutron star
(Strohmayer et al.  \cite{strohmayer96b}; Miller et al.
\cite{miller97}).  As $\nu_{\rm S} = \nu_{\rm K} - \nu_{\rm B}$, these
models predict that the difference in frequency $\Delta \nu$ of the twin
kHz peaks should remain constant, although $\nu_{\rm K}$ and $\nu_{\rm
B}$ may vary in time.  Obviously, the neutron star spin can not change
by 26\,\% in 3 days, as would be required by our observed change in
$\Delta \nu$.

In the case of Sco~X--1, where $\Delta \nu$ also varies (van der Klis et
al.  \cite{vanderklisetal97b}), it has been argued that the variations
can be attributed to near-Eddington accretion.  White \& Zhang
(\cite{white97}) propose a 35\,\% expansion of the neutron star
photosphere with conservation of angular momentum; Lamb (1996, private
communication) suggests that in near-Eddington accretion the height of
the inner disk increases, and that the different values of $\nu_{\rm K}
- \nu_{\rm B}$ are due to the different values of $\nu_{\rm K}$ at
different heights in the disk.  These explanations can not apply to
4U~1608--52, as at 3.6~kpc (Nakamura et al.  \cite{nakamura89}) its
luminosity was $1.3 \times 10^{37}$~erg~s$^{-1}$ and $9.4 \times
10^{36}$~erg~s$^{-1}$ on March 3 and March 6, respectively, less than
10\,\% of $L_{\rm Edd}$.

One possible way out might be time-variable scattering.  In the
beat-frequency model the lower frequency peak is expected to be at least
as broad as the higher frequency one, as it is generated by a beat
between the higher frequency QPO and the (coherent) neutron star spin.
Both on March 3 and 6 we observe the opposite.  This might be due, in
our shift and average technique, to a change in $\Delta \nu$ during each
observation.  Alternatively, a rapidly variable scattering medium around
the neutron star could cause the broadening, and perhaps even a
frequency shift.  The QPO at $\nu_{\rm K}$ (which in the sonic model is
a beaming oscillation), would be more sensitive to this than the
(luminosity oscillation) peak at $\nu_{\rm B}$.  It remains to be seen
if scattering can explain all the phenomenology:  the varying $\Delta
\nu$, the different peak widths at $\nu_{\rm K}$ and $\nu_{\rm B}$, the
energy dependences of the peak widths and amplitudes, and the
$\sim$27$\mu$s time lags between the high and low energy photons in the
lower frequency peak (Vaughan et al.  \cite{vaughan97}).

We therefore reconsider the idea of a layer at the surface of the
neutron star that does not corotate with the body of the star, but has
its own rotation frequency $\nu_{\rm L} \geq \nu_{\rm S}$, and that it
is this frequency $\nu_{\rm L}$ that beats with $\nu_{\rm K}$ (White \&
Zhang \cite{white97}).

The observations of frequency drifts in burst QPO (Strohmayer et al.
\cite{strohmayer97}) strongly suggest that such non-corotating layers,
at least at some times, exist.  However, the situation in 4U~1608--52
(and Sco~X--1) differs from that in X-ray bursts, in that the time
scales over which particular values of $\nu_{\rm L}$ persist are hours
to days, not seconds as in bursts.  Therefore, in these cases, $\nu_{\rm
L}$ can not be estimated by only considering changes in the moment of
inertia $I_{\rm L}$ of the layer as proposed by White \& Zhang
(\cite{white97}) for Sco~X--1, but must involve an estimate of, and may
well be dominated by, the changes in its angular momentum, $J_{\rm L}$.
The layer will gain angular momentum from the accreting matter, which
comes in with a higher angular velocity than that of the layer, and it
will lose angular momentum through friction with the underlying neutron
star body.  The layer will rotate differentially, and it will depend on
the precise nature of the beat frequency interaction which frequency is
picked out to interact with the Keplerian frequency.  If the
beat-frequency interaction takes place via beamed emission from a
certain radius in this layer out to the Kepler radius, such as for
example in the sonic point model (Miller et al.  \cite{miller97}), then
the photospheric radius will be the radius whose frequency we see.

Now consider the situation of 4U~1608--52 in our observations.  Previous
to the outburst there was little accretion for a considerable amount of
time, and the layer may well have settled down to near-corotation with
the body of the star.  The onset of the outburst produces a sudden
increase in $\dot M$, and high angular momentum matter is deposited in
the existing layer which consequently spins up.  Ignoring friction we
find that for a total mass of the layer of order $10^{-9}$M$_{\odot}$,
we can reproduce the observed frequency changes.

We note, that in Sco~X--1 it is observed that $\nu_{\rm L}$ decreases
when $\dot M$ increases, and vice versa.  Large changes in radius, and
therefore moment of inertia of the layer, with $\dot M$, as proposed by
White \& Zhang (\cite{white97}), would initially change $\nu_{\rm L}$ in
the sense observed, but it depends on the balance between angular
momentum gained in accretion and lost to friction what would happen
next.  In general, we would expect in such scenario that, after its
initial rapid spin down in response to an $\dot M$-induced increase in
$R$, the layer would gradually spin up again due to the angular momentum
deposited in it by the accreting matter, up to the point where friction
and accretion are in equilibrium.  It remains to be seen if the data
confirm this.

Finally, we note that the beat-frequency interpretation requires some
kind of pattern on the surface of the star that can interact with the
Keplerian blobs.  As for a non-corotating layer this pattern can not be
attached to the magnetic field lines (in fact, such layers would be
likely to smear out the hot spots at the magnetic poles into rings
around the star, and therefore could explain the absence of strong
pulsations in the persistent emission) some other origin is required for
it.  Perhaps magnetic loops or turbulent cells (prominences or
granulation) could fulfill this role.

\acknowledgements

This work was supported in part by the Netherlands Organization for
Scientific Research (NWO) under grant PGS 78-277 and by the Netherlands
Foundation for research in astronomy (ASTRON) under grant 781-76-017.
MM is a fellow of the Consejo Nacional de Investigaciones
Cient\'{\i}ficas y T\'ecnicas de la Rep\'ublica Argentina.  WHGL
acknowledges support from the National Aeronautics and Space
Administration.  JVP acknowledges support from the National Aeronautics
and Space Administration through contract NAG5-3269.  FKL acknowledges
support from NSF, through grant AST 93-15133, and NASA, through grant
5-2925.

\clearpage

\begin{deluxetable}{lrrrrrrr}
\scriptsize
\tablecolumns{8}
\tablecaption{Properties of the kHz QPOs in 4U~1608--52
\label{tabpow}
}
\tablewidth{0pt}
\tablehead{
\colhead{}                                          &
\multicolumn{2}{c}{Lower frequency peak}            &
\colhead{}                                          &
\multicolumn{2}{c}{Higher frequency peak}           &
\colhead{}                                          \nl
\cline{2-3} \cline{5-6}                             \nl
\colhead{Start time (UTC)/ Exposure [s]}            &
\colhead{rms [\%]}                                  &
\colhead{FWHM [Hz]}                                 &
\colhead{}                                          &
\colhead{rms [\%]}                                  &
\colhead{FWHM [Hz]}                                 &
\colhead{$\Delta \nu$ [Hz] \tablenotemark{a}}       &
\colhead{$\chi^{2}_{\nu}$/d.o.f. \tablenotemark{b}} \nl
}
\startdata
1996 Mar 3 19:18:24/ 6784                  & $7.37 \pm 0.04$ 
 & $4.83 \pm 0.08$ & & $\phm{0}3.14 \pm 0.40$ & $100^{+ 43}_{-32}$
 & $232.7 \pm \phm{0}6.6$ & 1.04/2229 \nl
1996 Mar 6 03:20:32/ 8320                  & $7.93 \pm 0.04$
 & $4.72 \pm 0.07$ & & $\phm{0}2.82 \pm 0.37$ & $ 49^{+ 23}_{-16}$
 & $293.1 \pm 11.5$       & 1.08/2227 \nl
1996 Mar 9 18:02:56/ 5696\tablenotemark{c} & $< 4 - < 7$\phm{0}\tablenotemark{d}
 & \nodata         & & $15.31 \pm 0.54$       & $129^{+ 13}_{-12}$
 & \nodata                & 1.02/2227 \nl
\enddata
\tablenotetext{a}{$\Delta \nu$ is separation between both peaks. As the
QPO peaks were arbitrarily shifted (see text), we do not quote their
centroid frequencies.}
\tablenotetext{b}{Reduced $\chi^2$ and degrees of freedom of the fit.}
\tablenotetext{c}{Only one QPO is seen. From our data we can not establish
if it is the upper or the lower peak.}
\tablenotetext{d}{95\,\% confidence, for FWHM 10 and 200~Hz, respectively.}
\tablenotetext{}{Quoted errors represent 1$\sigma$ confidence intervals.}

\end{deluxetable}


\clearpage

\onecolumn

\begin{figure}[h]
\plotfiddle{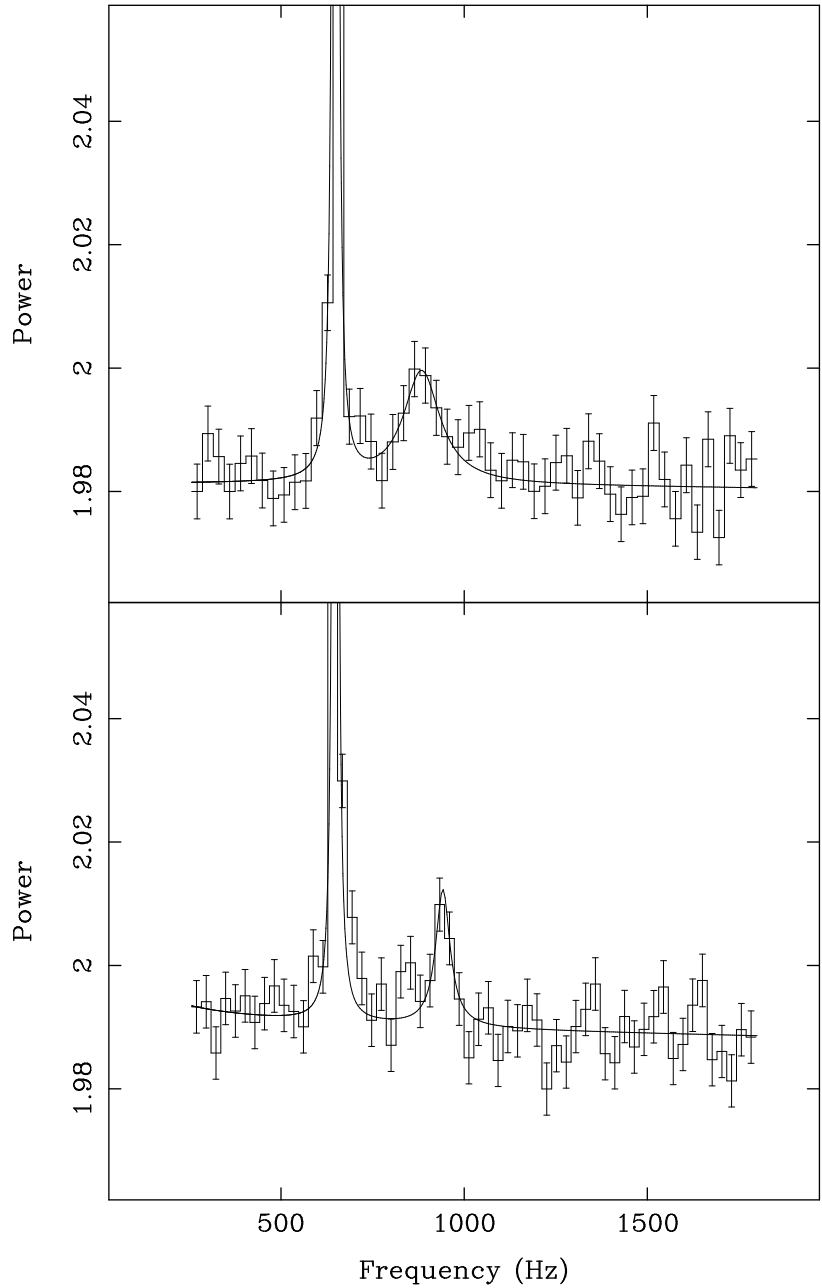}{250pt}{0}{70}{70}{-200}{-230}
\vspace{7.5cm}
\caption{
Power spectra for the observations of 1996 March
3 (upper panel) and March 6 (lower panel).  The frequency of the
strong peak was arbitrarily shifted to be the same in both power
spectra (see text for details).  On March 3 the data cover the full PCA
energy band (2 -- 60 keV), while on March 6 only data from 2 to 12.7~keV
were available.
\label{psfig}}
\end{figure}

\clearpage

\begin{figure}[h]
\plotfiddle{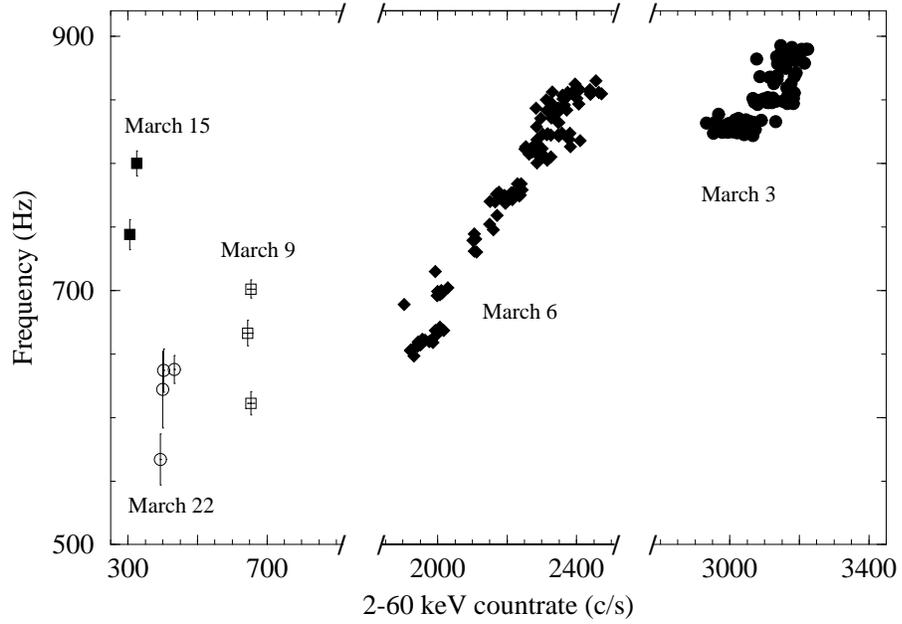}{250pt}{270}{70}{70}{-230}{400}
\caption{The QPO frequency vs.  count rate.  The data for
March 3, 6, 9, 15, and 22 are plotted using filled circles, filled
diamonds, open squares, filled squares and open circles,
respectively.  The points for March 15 and 22 are from Yu et al.
1997.  \label{freq_rate}}
\end{figure}

\end{document}